# Measuring the right thing: justifying metrics in AI impact assessments

*Stefan Buijsman and Herman Veluwenkamp*

## 1. Introduction

In order for us to use AI responsibly it is important that we can assess whether an AI system has the intended effects, and does not cause unforeseen harm. Does it indeed meet the various criteria that we have for a responsible or trustworthy AI system, for example as proposed by the European High Level Expert Group's Guidelines for Trustworthy AI?[1] Likewise, is a system that is advertised as being sustainable on the basis of a set of metrics about its environmental impact indeed actually sustainable? There is a general question here of how we know that the metrics that we are using to estimate the impact of an AI system are the right ones. This is separate from whether we can indeed produce reliable numbers for the metrics. Rather, it is about knowing whether we measure the right thing in the first place. After all, it frequently happens that an AI system that was developed with good intentions ends up having an impact that we ultimately find problematic. Often, these impacts are missed because we weren't looking in the right place.

This has frequently happened with the impact of AI systems. For example, it took several years before the biases in facial recognition systems (which were far less accurate for faces of people with darker skin tones than for faces of people with light skin tones) were noticed. Similarly, Obermeyer et al found that an algorithm that had been in use from at least 2013 onwards was heavily biased against patients who self-identified as black.[2] These patients got lower risk scores for the same symptoms, and as a result were offered less medical assistance than equally ill white patients. Likewise, the impact of navigation algorithms such as Google Maps on congestion in local roads and (relatedly) traffic safety took years to be addressed.[3] At that point, the optimization of people's individual journey times had already caused an increase in average journey time as well as a large spike in residential traffic, as navigation apps reroute traffic via local roads to avoid delays on the highway. To avoid cases such as these, impact assessments and audits are needed and indeed coming to the fore. However, for effective assessments and audits we need to know what metrics are the right ones to include.

That question is the focus of this chapter. We first motivate (in section 2) that there is a real need to justify one's choice of metrics in (AI) impact assessments. Strategies such as including all possible metrics for e.g. fairness (seen in for example the Model Cards proposal for reporting on AI systems)[4] do not suffice to get around the issue of selecting metrics. Nor do we have good ways to make these choices on a purely technical basis. It requires conceptual work to justify one's choice of metrics. We then introduce a framework, drawing on philosophical work from conceptual engineering, on how one can justify the choice of metrics through a two-step procedure from concepts to conceptions and from conceptions to metrics. To illustrate how this can work in practice we look at the challenge of finding metrics for ethical values, and specifically, fairness, in section 4. There we show how choices between different fairness metrics can be justified. This will give an approach that can help us argue for and evaluate the choice of metrics in impact assessments and audits.

---

[1] Smuha, N. A. (2019). The EU approach to ethics guidelines for trustworthy artificial intelligence. *Computer Law Review International*, *20*(4), 97-106.
[2] Obermeyer, Z., Powers, B., Vogeli, C., & Mullainathan, S. (2019). Dissecting racial bias in an algorithm used to manage the health of populations. Science, 366(6464), 447-453.
[3] Kulynych, B., Overdorf, R., Troncoso, C., & Gürses, S. (2020, January). POTs: protective optimization technologies. In *Proceedings of the 2020 Conference on Fairness, Accountability, and Transparency* (pp. 177-188).
[4] Mitchell, M., Wu, S., Zaldivar, A., Barnes, P., Vasserman, L., Hutchinson, B., ... & Gebru, T. (2019, January). Model cards for model reporting. In *Proceedings of the conference on fairness, accountability, and transparency* (pp. 220-229).

## 2. Metrics need to be justified

We need such an approach to justifying metrics, because these are difficult choices with serious consequences. It is a challenge, in particular, to make these choices in a systematic and well-motivated manner. This difficulty can most easily be seen in the case of ethical impacts. These are already the subject of AI audits, as in ethics-based auditing[5] which covers the entire range of AI development: ``functionality audits focus on the rationale behind decisions; code audits entail reviewing the source code of an algorithm; and impact audits investigate the types, severity, and prevalence of effects of an algorithm's outputs."[6] At each of these stages an assessment has to look at ethical values such as fairness, explainability, autonomy, and meaningful human control. But, how does one measure whether something is ethical?

To take up the example of fairness, there is a wide range of (statistical) fairness measures available that can be measured given the relevant data on sensitive variables.[7] However, there is a wide range of conflicting statistical (and other, such as causal) measures available and by now a number of mathematical impossibility theorems have shown that in virtually all cases it is mathematically impossible to optimize for all of these measures at the same time.[8] Picking one metric from among the many available is no easy task. Yet, we have to choose in our assessments between these metrics. Even though suggested internal documents such as model cards[9] would list a wide range of accuracies across groups as well as statistical fairness measures such an exhaustive treatment is insufficient. Once we have all these numbers available we, namely, have to decide whether we find these impacts acceptable. To do so, we need to know which fairness metric is the most relevant to the particular application in question.

For example, in an application where an AI system decides whether an applicant receives a loan or not there are different fairness metrics that we may measure. One can look at demographic parity, the rate at which different groups received loans. Alternatively, one could compare true positive rates across groups or false positive rates, just to name three of the many options available. As the impossibility theorems tell us, we cannot optimize all three of these at the same time. So, while an all-encompassing impact assessment may give us a wide set of fairness metrics to consider, there is still the question of which of these metrics we should pay the most attention to. Is it better to have very similar loan approval rates, or is it preferable to have equal false positive rates? Behind these choices are very real impacts and in the case of loan approvals these have been investigated in long-term simulation models. Liu et al show that optimizing for demographic parity (in addition to accuracy) can cause active harms to underprivileged groups.[10] In order to equalize the selection rate these groups are given a large number of loans that in fact they cannot afford, causing them to default. As defaults are expensive, and worsen people's financial position considerably compared to when they had not gotten a loan, this harm can (in a substantial number of the simulations run by Liu et al) be worse than the benefit of giving more people access to loans. If we, instead, optimize the false positive rates so that these are equal (and ideally low in absolute terms, as we also optimize the accuracy of the system) across groups then this disparity in defaults does not occur.

---

[5] Mökander, J., & Floridi, L. (2023). Operationalising AI governance through ethics-based auditing: an industry case study. AI and Ethics, 3(2), 451-468.
[6] Mökander, J., & Floridi, L. (2021). Ethics-based auditing to develop trustworthy AI. *Minds and Machines*, *31*(2), 324.
[7] Carey, A., & Wu, X. (2022). The statistical fairness field guide: perspectives from social and formal sciences. AI and Ethics, 1–23 & Mehrabi, N., Morstatter, F., Saxena, N., Lerman, K., & Galstyan, A. (2021). A survey on bias and fairness in machine learning. ACM Computing Surveys (CSUR) 54(6), 1–35.
[8] Kleinberg, J., Mullainathan, S., & Raghavan, M. (2016). Inherent trade-offs in the fair determination of risk scores. arXiv preprint arXiv:1609.0580.
[9] Mitchell, M., Wu, S., Zaldivar, A., Barnes, P., Vasserman, L., Hutchinson, B., ... & Gebru, T. (2019, January). Model cards for model reporting. In *Proceedings of the conference on fairness, accountability, and transparency* (pp. 220-229).

It's tempting now to say that the fairness metrics were simply the wrong metric to look at. After all, the difference above is one in terms of people's financial position, and that is an impact we can estimate independently of the different (statistical) comparisons captured in fairness metrics. There is a pragmatic reason not to do so, as we have a range of methods to include statistical fairness metrics in the training procedure of AI systems. On the other hand, it will be more difficult to include a criterion to maximize the absolute impact on people's financial position in that procedure. On a more principled basis, we can see statistical distributions of AI outputs as one of the relevant factors that impacts people's long-term financial position. By keeping track of this statistical distribution, we can separate that impact from other aspects of the AI system that impacts people's financial position. The question of what statistical metric is the most relevant to capture the fairness of the system may then be approached from the standpoint of what metric has the most impact on people's financial situation, but that does not mean that we should do without fairness metrics altogether.

At the same time, it is far from clear that these statistical metrics suffice to capture the fairness impacts of an AI system. For example, they need not give sufficient insights into whether algorithms are used responsibly at an organisational level.[11] There are metrics developed for organizational audits that may help here, as well as work on KPIs for Responsible Research and Innovation.[12] But even when we include these in the impact assessment of an AI system, there is a debate on whether fairness metrics actually capture the substantive notion of (distributive) justice that we are after[13] and there is a range of papers that have criticized fairness metrics for falling short of the goal to measure whether a system is just and fair.[14] It is thus very much an open question what we should measure in order to assess the ethical impacts of an AI system, and in fact Lacroix and Luccioni have argued that it is impossible to develop benchmarks for how ethical an AI system is.[15]

This challenge is reflected in current methods that aim to assess the impacts of AI systems. Ethics-based auditing gives us no clear guidance here, and other methods such as Z-Inspection, discussed in chapter

---

[10] Liu, L., Dean, S., Rolf, E., Simchowitz, M., & Hardt, M. (2018). Delayed impact of fair machine learning. In International Conference on Machine Learning. PMLR, 3150–3158.
[11] Minkkinen, M., Niukkanen, A., & Mäntymäki, M. (2022). What about investors? ESG analyses as tools for ethics-based AI auditing. AI & Society, 1-15.
[12] Kwee, Z., Yaghmaei, E., & Flipse, S. (2021). Responsible research and innovation in practice an exploratory assessment of Key Performance Indicators (KPIs) in a Nanomedicine Project. Journal of Responsible Technology, 5, 100008.
[13] Kuppler, M., Kern, C., Bach, R. L., & Kreuter, F. (2021). Distributive justice and fairness metrics in automated decision-making: How much overlap is there?. arXiv preprint arXiv:2105.01441.
[14] E.g., Selbst, A. D., Boyd, D., Friedler, S. A., Venkatasubramanian, S., & Vertesi, J. (2019). Fairness and abstraction in sociotechnical systems. In Proceedings of the conference on fairness, accountability, and transparency (pp. 59-68).
[15] LaCroix, T., & Luccioni, A. S. (2022). Metaethical Perspectives on'Benchmarking'AI Ethics. arXiv preprint arXiv:2204.05151.

X in this volume, rely on expert judgement. Z-inspection takes the EU High-Level Expert Group values as its starting point,[16] and then faces the challenge on how to measure the impacts of an AI system in light of these values. As can be seen, for example, in a case study for the framework involving an algorithm that recognizes cardiac arrests in emergency calls[17] this framework then proceeds from a wide identification of stakeholders and their values to the analysis of (socio-)technical scenarios to reach an identification and (potentially) resolution of ethical, technical and legal issues of an AI system. The idea is that a large and interdisciplinary group of experts can help to make the translation from these values into specific metrics, but the question remains what the choices of these experts (should be) based on and how they might justify their choice.

These difficulties are there for ethical impacts such as those related to fairness, but they are just as present for other impacts that we aim to measure. What are the right metrics to use for sustainability, for example? A good example here is the concern that companies engage in greenwashing, which (as de Freitas Netto et al discuss in a review of the topic) is to a large extent characterized by selective disclosure of information – performance on metrics – about environmental performance.[18] We find this problematic precisely because those metrics are often not the (only) ones that we care about. For example, Delmas and Burbano define greenwashing as "poor environmental performance and positive communication about environmental performance".[19] This shows that when the company had chosen the *right* metrics the results would have been worse than on the metrics that they in fact chose. The choice of metrics is, in other words, not properly justified, because it fails to capture their actual sustainability.

Social impacts are often no easier to measure. Safety, for example, is difficult to capture in a set of metrics. There are many kinds of safety at stake, as there are many types of risks that may impact the overall value of safety. Not only that, not all types of safety can be easily quantified. Mental safety, for example, will be far harder to quantify than one's risk of dying from a particular disease (where we can look at survival rates). Here, too, then we run into the challenge of ensuring that we pick the right metric to truly capture the overarching value that we are interested in. And the list goes on, as health, well-being and many other values that we find important are difficult to capture in the metrics we need for impact assessments. So how can we justify our choice of metric for these values that are difficult to assess? We suggest a stepwise procedure in the next section and illustrate how this may work in practice with two examples regarding the measurement of fairness in section 4.

## 3. Conceptual engineering: linking metrics to justified conceptions

To understand how we can justify different metrics for a concept, it is helpful to look at methods developed in the field of conceptual engineering. Conceptual engineering is a sub-discipline within analytic philosophy that focuses on the critical evaluation and subsequent improvement of our words and concepts. It is driven by the need for beneficial changes, whether social, theoretical, or political. Conceptual engineering finds its applications in diverse areas such as redefining "planet" by the International Astronomical Union, changing the content of "marriage" in the context of shaping law and in social policy when we investigate how best to understand a term like "woman". Far from being a mere academic exercise, it's a discipline that often operates across various sectors.

---

[16] Zicari, R. V., Brodersen, J., Brusseau, J., Düdder, B., Eichhorn, T., Ivanov, T., ... & Westerlund, M. (2021). Z-Inspection®: a process to assess trustworthy AI. IEEE Transactions on Technology and Society, 2(2), 83-97.
[17] Zicari, R. V., Brusseau, J., Blomberg, S. N., Christensen, H. C., Coffee, M., Ganapini, M. B., ... & Kararigas, G. (2021a). On assessing trustworthy AI in healthcare. Machine learning as a supportive tool to recognize cardiac arrest in emergency calls. Frontiers in Human Dynamics, 3, 673104.
[18] de Freitas Netto, S. V., Sobral, M. F. F., Ribeiro, A. R. B., & Soares, G. R. D. L. (2020). Concepts and forms of greenwashing: A systematic review. Environmental Sciences Europe, 32(1), 1-12.
[19] Delmas M, Burbano V (2011) The drivers of greenwashing. Calif Manag Rev 54(1):67.

Conceptual engineering is relevant in the context of our paper. When we are determining what the best metric for a concept is, we are also engaging in conceptual engineering.[20] This is exemplified in ethical machine learning, where the aim is not just to apply fairness, but to identify the most suitable way to quantify it within mathematical models. The process of choosing these metrics actually forms an essential part of conceptual engineering itself, guiding us to more refined or reimagined criteria for what should constitute fairness.

An important distinction, which we will employ as well in this paper, is the distinction between "concepts" and "conceptions". A concept is a generalized, abstract representation of the values we aim to measure, whereas conceptions are specific interpretations of these broader ideas. The relationship between concepts and conceptions is one-to-many; for instance, "justice" is a broad concept that may include more concrete conceptions like "retributive justice", "social justice" and "procedural justice".

Conceptual Engineering has its theoretical roots in the works of philosophers like Carnap[21] and in metaphilosophy,[22] and it has recently converged with parallel efforts in social philosophy under the term "ameliorative analysis".[23] Conceptual engineering is structured around two main pillars: the theoretical and the practical.[24] On the theoretical side, researchers explore foundational issues such as how this approach should operate on concepts or expressions,[25] how it integrates with theories of mind and language,[26] and how its proposed changes can be effectively implemented.[27] On the practical side, the focus is on the specific concepts and their conceptions that need revising. For example, one might examine the concept of "truth" and propose more nuanced conceptions to better handle the complexities inherent in natural language semantics.[28]

To operationalize these theoretical and practical pursuits, the field outlines methods for effective conceptual engineering, including the identification of specific processes, target concepts and their conceptions, and overarching goals that guide the various projects within this discipline.[29] Whether one is interested in refining social conceptions related to justice[30] or tweaking scientific categories for better clarity,[31] conceptual engineering offers a systematic way to carry out these changes.

---

[20] Veluwenkamp, H., & van den Hoven, J. (2023). Design for values and conceptual engineering. *Ethics and Information Technology*, *25*(1), 2. https://doi.org/10.1007/s10676-022-09675-6

[21] Carnap, R. (1950). *Logical foundations of probability* (Vol. 2). University of Chicago Press.

[22] Eklund, M. (2021). Conceptual Engineering in Philosophy. In J. Khoo & R. Sterken (Eds.), *The Routledge Handbook of Social and Political Philosophy of Language*.

[23] Haslanger, S. (2012). *Resisting Reality: Social Construction and Social Critique*. Oxford University Press. https://doi.org/10.1093/acprof:oso/9780199892631.001.0001.

[24] Isaac, M. G., Koch, S., & Nefdt, R. (2022). Conceptual engineering: A road map to practice. *Philosophy Compass*, e12879.

[25] Cappelen, H. (2018). *Fixing language: An essay on conceptual engineering*. Oxford University Press.

[26] Löhr, G., & Michel, C. (2023). Conceptual engineering, predictive processing, and a new implementation problem. *Mind & Language*.

[27] Nimtz, C. (2021). Engineering concepts by engineering social norms: Solving the implementation challenge. *Inquiry*, 1–28.

[28] Scharp, K. (2013). *Replacing Truth*. Oxford University Press UK.

[29] Thomasson, A. L. (2020). Pragmatic Method for Normative Conceptual Work. In A. Burgess, H.

With this understanding of the theoretical and practical dimensions of conceptual engineering in place, we now turn to a specialized application of these principles: a two-phase framework for justifying the selection of a specific metric. This framework serves as a systematic approach to enact the principles of conceptual engineering within the specific scope of metric selection.

The first phase of the framework focuses on choosing an appropriate conception for the concept in question. The question which conception is justified is context-dependent. For instance, choosing a conception of "fairness" in a legal setting might prioritize procedural consistency and equality before the law, whereas in a healthcare setting, the focus might be on equitable access to resources and services. The chosen conception serves as the foundation upon which we will later develop a quantifiable metric. This makes it crucial to ensure that it aligns well with the contextual demands and ethical imperatives of the situation.

After determining the appropriate conception, the next phase is to methodically transform it into a measurable metric. This involves a detailed analysis to identify the core attributes that make up the larger concept. For instance, if the chosen conception emphasizes 'equitable distribution,' then that serves as the normative focus when devising a metric. Subsequently, these attributes are weighted to create a composite metric that aligns closely with both the context and the normative function of the concept. This ensures that the metric not only quantifiably measures but also ethically embodies the selected conception.

## 3.1 From concept to conception

Having laid out the overarching structure of our two-phase framework, we now present each phase in greater detail. An increasingly popular methodology in conceptual engineering holds that the *function* of a concept determines how we should engineer them.[32] This entails that we shouldn't merely look at the representational qualities of our concept, but that we should consider that concepts can do many other things for us. The first step, therefore, involves identifying why it's valuable to have the concept in question. Let us call this the *normative* function of the concept.[33] For "fairness", arguably the most important value in having this concept in our repertoire is to guide human interactions, promoting social harmony and trust.

It is important to note that the normative function needn't correspond to the *actual* function of a concept. This is so, because the actual function might be morally problematic. For instance, consider the concept of "justice" in a society where retributive measures are disproportionately applied against certain marginalized groups. Here, the actual function of "justice" within that societal system may perpetuate inequality and systemic bias. However, the normative function, based on ethical deliberation, might aim to correct these biases by emphasizing equality and fairness. In this case, the actual function conflicts with the normative function. This highlights the critical role of normative deliberation in refining or reimagining a concept to better align with ethical imperatives and societal goals.

---

[30] Podosky, P.-M. C. (2018). Ideology and normativity: Constraints on conceptual engineering. *Inquiry*, 1–15. https://doi.org/10.1080/0020174X.2018.1562374
[31] Simion, M. (2018). The 'should' in conceptual engineering. *Inquiry: An Interdisciplinary Journal of Philosophy*, *61*(8), 914–928.
[32] Thomasson, A. L. (2020). Pragmatic Method for Normative Conceptual Work. In A. Burgess, H.; Simion, M., & Kelp, C. (2020). Conceptual Innovation, Function First. *Noûs*, *54*(4), 985–1002; Queloz, M. (2021). *The practical origins of ideas: Genealogy as conceptual reverse-engineering*. Oxford University Press.
[33] Köhler, S., and Veluwenkamp, H. (2024). "Conceptual Engineering: For What Matters." *Mind*.

The normative function of a concept serves as a practical tool for justifying the use of a specific conception in a given context. Once the function is determined, it can act as a criterion against which various conceptions can be evaluated. The idea is to select the conception that best fulfills the identified function within the specific context at hand. In essence, the 'best' conception is the one that aligns closely with the normative function, enabling it to address the relevant ethical, social, or legal considerations.

## 3.2 From conception to metrics

This brings us to the second phase, which involves translating the chosen conception into a quantifiable metric. The first step in this phase is to recognize the normative significance of the selected conception. The conception will outline what is normatively important about the concept in the given context.

The next step is to dissect the selected conception into discrete, measurable elements. This involves a detailed analysis to identify the core attributes that make up the larger concept. For example, if the conception focuses on "equitable distribution", one could consider attributes like income level, access to healthcare, or educational opportunities as measurable elements. It's imperative at this stage to ensure that these elements are justifiable within the scope of the specific context and the normative function of the concept. For instance, one should ask: Do these elements collectively capture the essence of 'equitable distribution' in a particular setting? If the context is healthcare, income level might be less relevant than access to medical resources, thereby requiring justification for its inclusion or exclusion.

Finally, the metric is formally defined, but not before its validity and relevance are systematically justified. In this stage, it's important to determine the weighting of each attribute to create a composite metric. Again, the justification for the weighting should be transparent and backed by empirical evidence and/or reasoned argumentation. For example, in the case of "fairness," one might have to decide whether income equality should be weighted more heavily than educational access in a given context. The justification for this weighting can be rooted in the normatively significant elements of the selected conception identified earlier. This ensures a coherent and ethical approach. Incorporating these steps for justification makes sure that the developed metric is not only quantifiable but is also normatively justified, aligning closely with the selected conception and its normative function.

## 3.3 From Concepts to Sub-Concepts

We cannot always follow the exact route sketched above. In some cases, we start off with a single concept such as fairness or justice, but end up looking at multiple concepts (distributive justice, procedural justice) each with their own conception. For example, when examining responsibility in the context of self-driving cars, the original concept of "responsibility" may not sufficiently capture all relevant nuances involved. We find ourselves needing to focus on multiple *sub-concepts*, each with their own individual conceptions and applicable metrics. In this scenario, our framework must be sufficiently flexible to accommodate this complexity. Take, for instance, the case of self-driving cars where the broader concept of 'responsibility' breaks down into "causal responsibility" and "moral responsibility". These sub-concepts each have unique conceptions which we are interested in measuring. "Causal responsibility" might focus on measurable factors like software reliability and sensor accuracy, with metrics perhaps derived from incident reports and failure rates. "Moral responsibility," on the other hand, would focus on ethical dimensions, which might be quantified through societal surveys on acceptable risk or ethical algorithms. Each sub-concept aligns with a different normative function: one oriented toward empirical measurability and predictability, and the other geared toward praise- and blameworthiness.

When is this the right way to go? That is, when should we split up a concept that we aim to measure into two sub-concepts, each associated with their own conception? At first, it may be tempting to focus on the application of the concepts. If we look at the example of responsibility, then there is a clear way in which the sub-concepts of causal and moral responsibility come apart. One can be morally responsible for the harm done by a self-driving car, without being directly causally implicated in the operation of the car. Vice versa, one can be causally responsible for a harm without being blameworthy for the harm.[34]

The mere fact that such a division can be made, however, cannot be all there is to it. It's often trivial to split concepts into sub-concepts by introducing extra distinctions, e.g. we can apply a-justice to all decision-making processes whose name starts with 'a' up to and including those starting with 'm' and b-justice to all decision-making processes whose names starts with 'n' up to and including those starting with 'z'. If some overlap is needed then we can simply adjust the range of the letters to ensure that there is some common ground, but also some differences between the two types of justice. Clearly, such a distinction is not helpful nor should it have any consequences for impact assessments. So, when does it make sense to divide up a concept, to work with more specific ones instead?

A more tempting idea is to look at whether the sub-concepts align with two distinct functions that need to be addressed. The challenge is to spell out precisely what this functional approach entails. One approach is to look at whether there are regularities that such a division would exploit. For example, consider the concept of water. The primary function of water in a biological context is to act as a solvent, medium for transport, and a key participant in various chemical reactions. Any additional details, like whether the water comes from a lake or a well, may not significantly contribute to this primary function. We can exploit this same idea to give a bit more of a handle on when splitting up concepts into sub-concepts is a good idea. If there are distinct functionalities associated with two phenomena that fall under the same concept then we can expect that it is no more complex to describe them separately as it is to describe them jointly. For example, although 'renates' and 'cordates' (creatures having a kidney and creatures having a heart, respectively) as a matter of fact describe the same set, each serves different biological functions. We would be forced into a description such as 'creatures having a heart and/or a kidney' (depending on one's preference for how to combine them). This would be just as long a description as when we have the two concepts separately, so the division follows a regularity that is out there, in the world. This clearly does not happen in the case of our toy example of a-justice and b-justice, since separating out decision making procedures by the first letter in their name does not follow any regularities out there: 'a-justice and b-justice' will not capture any more than plain 'justice'.

Looking at the distinction between distributive and procedural justice again we can see that the idea of looking at regularities also works in that case. Distributive justice is about the allocation of benefits and harms, procedural justice is about the decision-making procedures themselves. There is a natural way in which justice can be split up into these two parts, and so there is a regularity to follow in the split of the concepts; for justice in general we will have to think about both the process and the outcomes.

---

[34] Mecacci, G., & Santoni de Sio, F. (2020). Meaningful human control as reason-responsiveness: The case of dual-mode vehicles. *Ethics and Information Technology*, *22*(2), 103–115.

Still, if the two dimensions are intrinsically linked, as Rawls' theory[35] suggests for distributive and procedural justice, then maintaining a division may be unwarranted. If we follow Rawls then we have to think about the outcomes when we are evaluating whether a process is procedurally just, for example. Perhaps, in such a case, a single conception of justice in terms of achieving the right kinds of outcomes (often enough) would be better. There would be no need for the (in this scenario additional details of) inclusion of properties of the procedure itself. That, again, would be quite fitting: if it turns out that we need to know nothing about the procedure other than the kind of outcomes it leads to, then we only need to measure those outcomes. A separate conception of the procedural justice of the decision-making process is then superfluous. Since discussions are ongoing, and Rawls' account is controversial, it still makes sense to keep them separate. It is, nevertheless, a good illustration of how this choice of whether to split up a concept or not can be influenced by how we think about the concepts and thus about the world.

## 4. From concept to metric: fairness

One of the reasons we consider fairness so explicitly in this paper is that recent work on this concept[36] offers a clear example of how the translation from concept to conception and then to metrics can be done in practice. It illustrates both that conceptions of concepts are often at the right level of abstraction to establish a principled choice for a metric and that different conceptions may yield different choices. We'll consider both in order, and then discuss these examples in light of the method presented in the previous section.

To start with, Buijsman[37] focuses on fairness in terms of distributive justice, which in turn is understood via the conception of distributive justice presented by Rawls.[38] By choosing this account there is a decision on how to understand fairness, to be used to derive metrics. Of course, this choice of conception needs to be justified, so there should be a defense of the conception of justice/fairness that Rawls presents in terms of the normative function of the concept in the particular setting we are looking at. Buijsman offers no such defense, and merely appeals to the influence of Rawls on current debates of distributive justice. That being said, these discussions on what particular account of distributive justice best fits the normative function of the concept, are extensive and we will therefore leave that as it is, and focus on the translation from a conception of fairness in terms of distributive justice to a choice of metric. We thus start with the two principles of justice that make up Rawls' account:

> 1. Each person has the same indefeasible claim to a fully adequate scheme of equal basic liberties, which scheme is compatible with the same scheme of liberties for all; and
> 2. Social and economic inequalities are to satisfy two conditions: first, they are to be attached

---

[35] Rawls, J. (2001). Justice as fairness: A restatement. Harvard University Press.
[36] Buijsman, S. (2023). Navigating fairness measures and trade-offs. AI and Ethics, 1-12 & Baumann, J., & Loi, M. (2023). Fairness and Risk: An Ethical Argument for a Group Fairness Definition Insurers Can Use. Philosophy & Technology, 36(3), 45.
[37] Buijsman, S. (2023). Navigating fairness measures and trade-offs. AI and Ethics, 1-12.
[38] Rawls, J. (2001). Justice as fairness: A restatement. Harvard University Press.

> to offices and positions open to all under conditions of fair equality of opportunity; and second, they are to be to the greatest benefit of the least-advantaged members of society.[39]

The first principle is less applicable to the kinds of algorithms that are typically under consideration, such as those supporting hiring decisions or determining insurance premiums. Instead, the second part is the most salient for these discussions of algorithmic fairness. There, Rawls presents us with the following picture: we should strive for equality of opportunity, such as equal access to education and healthcare, to provide as level a playing field as we can. However, some inequalities are permissible, as long as these inequalities are of net benefit to the worst-off. This conception can then be used to make a choice between statistical fairness measures, given a specific context.

Although Rawls had a specific context in mind when formulating this conception, i.e., the structure of institutions in a liberal democracy, the applicability of his ideas extends beyond this initial scope. However, as we already presented the setting of a bank using an algorithm to grant or deny a loan in section 2 we will focus on that. Is it, in this case, more important to look at whether two salient groups (e.g. men and women) receive loans at the same rate or to compare the false/true positive rate? As mentioned, these options are incompatible[40] and have a real impact on people's financial position. Buijsman argues that if we follow Rawls then we should choose the metric that most closely tracks the actual impact on (in this case) the financial situation of the group that is worst off.[41] Since simulation studies[42] show that false positives, i.e., loans that are granted but ultimately cannot be paid back, are by far the most costly of the four different outcomes we should focus on the distribution of these costly false positives. Rawls' account of course isn't concerned with making these perfectly equal, but rather with the absolute level of welfare of the marginalized group. Still, as optimizing for other fairness metrics was shown to cause harm precisely because of an increased disparity in false positives, it is very likely that by optimizing on equal false positive rates the absolute level of welfare is also optimized. If not, then there is a clear metric that can be used to determine an even fairer version of the algorithm (on this conception of fairness) by maximizing the financial position of the marginalized group.

We thus get a specific outcome for the bank case, that can be traced back to a motivated conception of fairness. There is, in short, a reason that we can appeal to in support of our choice of metric: it tracks the most costly type of case for applicants. For loan applications that's false positives, but in other settings it will be something else. Buijsman suggests that in the setting of healthcare diagnoses it may well be that false negatives are by far the most costly, because getting a true positive (and thus the correct treatment) is particularly important in healthcare.[43] This will, again, vary depending on the type of diagnosis that is at issue, as well as the controls in place to mitigate the harms of false positives, but it illustrates how the choice of metric may vary across application domains.

Even more striking are the differences between situations where Rawls' principle to maximize the welfare of the least-advantaged group applies, and those where it is trumped by the demand for fair equality of opportunity. In these settings, most notably those where hiring decisions are made, there is a clear instruction to arrange for equality between individuals. This is typically interpreted in causal terms and entails, among other things, that one's socio-economic status should have no impact on one's chances at a position.[44] Causal fairness metrics are of course available,[45] and so we have again a clear choice for a metric in these situations.

---

[39] Rawls, J. (2001). Justice as fairness: A restatement. Harvard University Press, pp. 42-43.
[40] Kleinberg, J., Mullainathan, S., & Raghavan, M. (2016). Inherent trade-offs in the fair determination of risk scores. arXiv preprint arXiv:1609.0580.
[41] Buijsman, S. (2023). Navigating fairness measures and trade-offs. AI and Ethics, 1-12.
[42] Liu, L., Dean, S., Rolf, E., Simchowitz, M., & Hardt, M. (2018). Delayed impact of fair machine learning. In International Conference on Machine Learning. PMLR, 3150–3158.
[43] Buijsman, S. (2023). Navigating fairness measures and trade-offs. AI and Ethics, 1-12.

Before moving on to the translation made by Baumann and Loi,[46] there is a further point of interest to discuss, namely that one concept may link to many metrics. First, there may be more than one metric that follows from a single conception. Statistical metrics on the AI system itself will, for example, not capture everything about Rawlsian equality of opportunity. Such equality of opportunity also involves giving underprivileged groups additional support and arranging for appropriate education, etc. Maximizing the welfare of the least-advantaged groups likewise requires more than tailoring decision-making procedures; it also calls for social welfare programs, among other things, which influences the types of decisions that are being made. Such actions are not covered by the statistical fairness metrics that (also) need to be selected. In short, while we have been considering a single metric that follows from the conception of fairness, in reality there are likely a number of metrics capturing different aspects that are in this conception.

In addition, and as discussed in section 3.3, it may be necessary to split up a concept into relevant sub-concepts. In the examples above we focussed on fairness conceived as stemming from distributive justice, and developed that with a particular account of distributive justice. However, there are other aspects of justice that are also important. A hiring procedure that offers equality of opportunity because it decides on who gets a job through a lottery may satisfy demands of distributive justice and demographic parity but it feels far from fair. We also want the procedure through which jobs are distributed to meet certain criteria, such as making decisions based on the merits of individual applicants. This would fall under the concept of procedural fairness, as an addition to the concept of distributive fairness considered here. We may perhaps translate procedural fairness into a measurable requirement of sufficient accuracy relative to whether candidates that are selected are indeed performing well in the job. We've made that point for fairness, but want to stress that this will hold in general: a single concept may need to be refined into several sub-concepts, each of which may have a number of dimensions that will in turn be captured by a set of metrics. It would be wrong to expect that a concept can always, or even often, be captured by a single metric.

With these cautionary notes we can turn to the other example regarding fairness, by Baumann and Loi.[47] They focus specifically on AI systems that determine the premiums for insurance policies, and look at a conception of fairness tailored to that context. The idea there is to start with a general maxim from Aristotle to treat like cases alike. This is linked initially to the notion of actuarial fairness: your premium should be equal to your risk. However, in practice it is (so they argue) not feasible to determine an individual's risk of e.g. getting into a car accident. Instead, we need a more practical way of determining premiums, and of understanding fairness, which relies on the more coarse-grained data that insurers have regarding claims and statistical correlations between claims and group properties (e.g., young drivers filing more claims related to car crashes).

---

Baumann and Loi introduce two conceptions to do precisely that, based on the premise that the function of insurance is to spread costs over a larger group and thus offer a measure of solidarity between the insured. Since fairness is based on treating like cases alike, they then see two possible conceptions of fairness in relation to this idea of solidarity:

1. Risk solidarity: individuals within a group share risks equally
2. Chance solidarity: individuals within a group facing similar risks share the costs resulting from the chance that a risk materializes equally

The first type of solidarity is one that is often seen in the context of health insurance, where premiums may be fixed for a certain package of treatments regardless of the risks an individual runs. There is in those cases no personalization of the premium based on your actual state of health or lifestyle. Rather, we have decided to collectively share these risks to ensure adequate treatment for everyone. From this, a notion of fairness follows that simply holds that premiums should be fixed, or perhaps even better holds that insurance coverage should be maximal (since we aim to ensure access to healthcare for everyone). From a specific conception of fairness – risk solidarity – follows a clear instruction on how to measure how fair the premiums are. We should look, in this case, at the degree to which premiums differ and are affordable to everyone, period. There will, however, be very few types of insurance to which we apply risk solidarity.

The second type of solidarity applies to the vast majority of insurance schemes (car insurance being the standard example). There, it follows from the conception of chance solidarity that the statistical fairness measures known as independence/demographic parity (giving equal premiums to, e.g., men and women, on average) and separation (among individuals with the same outcomes equalize average premiums across groups) would be the wrong metrics to pick. In the case where men and women have, on average, different risks then equalizing premiums for these two groups will mean that we are violating the principle that the costs of someone's risk is shared equally. Instead, the group with lower average risk will be paying more for their risk than the group with higher average risk: demographic parity thus involves risk solidarity as opposed to chance solidarity as soon as the risk profiles of the two groups differ. A similar argument can be made for separation, so the more interesting point to turn to is the positive argument that chance solidarity directly leads us to the fairness metric known as sufficiency or predictive parity.

According to the sufficiency metric the expected damages should be equal for equal paid premiums, across two groups. So, if subgroup A1 pays premium x with expected damages y then subgroup B1 should pay the same premium as long as their expected damages are the same. If those groups are large enough, then the expected damages can be equated with the average observed damages, showing that we're focussing on the chance occurrences of damages here. Baumann and Loi show that this indeed neatly captures the conception of chance solidarity, and highlight that because of the difficulty to estimate expected damages we should not be too hasty in drawing conclusions from a disparity in premiums.

What matters for our points here, is that both of these papers highlight a trajectory from an overarching value to a (justified) conception and finally down to (a) metric(s). Baumann and Loi have the arc from the function of insurance (providing a kind of solidarity) in a particular context that justifies a conception of fairness based on treating like cases alike. From this conception of risk or chance solidarity we can then derive particular metrics that are well-motivated in the insurance context.

Buijsman has a similar type of reasoning for AI systems that distribute harms and benefits.[48] The conception of fairness that fits in this case is one that looks at what distributions are most just. Rawls has argued for his interpretation of distributive justice in terms of it being endorsed by people in a situation where they don't know what their position in society will be, effectively linking it to the function of fairness we highlighted in section 3 to promote social harmony.[49] From that justified conception there is then a translation down to metrics. We thus *can* pick metrics in a non-arbitrary way, accompanied by a clear explanation why it is the appropriate metric to look at.

## 5. Conclusion

How can we make sure that we are measuring the right thing in an impact assessment? We've presented a two-step approach to answering that question for the many different kinds of concepts that we want to formulate metrics for. As we've shown, metrics can be formulated on the basis of conceptions of concepts. Once we have a conception, so a clearly worked out understanding of what a concept means in a particular context, then we can justify the choice of metrics based on that. We illustrated that this is indeed feasible in practice by showing how to derive metrics for (distributive) fairness from two different conceptions, a Rawlsian conception for a wide range of contexts and a conception of fairness as solidarity in the insurance context.

At the same time, these two competing conceptions (at least in the insurance context) may lead to different outcomes. The least advantaged members of society may well be better off if we subsidize their risks to a larger extent than on pure chance solidarity, for example. It is thus important to be able to choose between these competing conceptions, otherwise we still have an issue of determining the right metrics for our impact assessments. For this the more theoretical discussion of how well a conception fits with the function of a concept comes in, and the literature on conceptual engineering offers more examples and methods on how to do this.[50] In practice, we expect that such discussion will remain difficult, however, and thus a pragmatic solution would be to be transparent about the conceptions one adopts in impact assessments and the motivation for choosing those conceptions. There are, after all, better and worse choices of metrics and so having a clear framework to justify one's choices will help to make this visible.

---

[48] Buijsman, S. (2023). Navigating fairness measures and trade-offs. AI and Ethics, 1-12.
[49] Rawls, J. (2001). Justice as fairness: A restatement. Harvard University Press.
[50] Cappelen, H. (2018). *Fixing language: An essay on conceptual engineering*. Oxford University Press; Haslanger, S. (2012). *Resisting Reality: Social Construction and Social Critique*. Oxford University Press; Machery, E. (2009). *Doing without concepts*. Oxford University Press; Chalmers, D. J. (2020). What is conceptual engineering and what should it be? *Inquiry*, 1–18.